\documentclass[showpacs]{revtex4}
\usepackage{psfig}
\usepackage{amsmath}
\usepackage{graphicx}

\begin{document}

\title{Fluctuation-Dissipation relations in Driven Granular Gases}

\author{A. Puglisi, A. Baldassarri and V. Loreto}

\affiliation{``La Sapienza'' University in Rome, Physics Department,
P.le A. Moro 5, 00185 Rome, Italy and \\INFM, Center for Statistical
Mechanics and Complexity, Rome, Italy}

\date{\today}

\begin{abstract}
We study the dynamics of a 2d driven inelastic gas, by means of Direct
Simulation Monte Carlo (DSMC) techniques, i.e. under the assumption of
Molecular Chaos.  Under the effect of a uniform stochastic driving in
the form of a white noise plus a friction term, the gas is kept in a
non-equilibrium Steady State characterized by fractal density
correlations and non-Gaussian distributions of velocities; the mean
squared velocity, that is the so-called {\em granular temperature}, is
lower than the bath temperature.  We observe that a modified form of
the Kubo relation, which relates the autocorrelation and the linear
response for the dynamics of a system {\em at equilibrium}, still
holds for the off-equilibrium, though stationary, dynamics of the
systems under investigation.  Interestingly, the only needed
modification to the equilibrium Kubo relation is the replacement of
the equilibrium temperature with an effective temperature, which
results equal to the global granular temperature.  We present two
independent numerical experiment, i.e. two different observables are
studied: (a) the staggered density current, whose response to an
impulsive shear is proportional to its autocorrelation in the
unperturbed system and (b) the response of a tracer to a small
constant force, switched on at time $t_w$, which is proportional to
the mean-square displacement in the unperturbed system. Both measures
confirm the validity of Kubo's formula, provided that the granular
temperature is used as the proportionality factor between response and
autocorrelation, at least for not too large inelasticities.
\end{abstract}

\pacs{05.40.-a,45.70.-n,05.70.-a}
       
\maketitle


\section{Introduction}

In the past few years granular materials~\cite{reviews} have become a
fast growing field of research. In this framework, a strong interest
has arisen in the subject of the so-called granular gases, systems of
grains at very low density so that the collisions can be considered
always binary and instantaneous~\cite{luding}.  The collisions among
the grains are dissipative, and the amount of energy lost in each
collision is parametrized by the so-called restitution coefficient $r$
(see below for a precise definition).  Since the energy of the system is
therefore not conserved an external forcing is usually applied to
obtain a dynamic stationary state.  For this kind of systems, the
analogy with perfect gases allows to introduce a {\em granular}
temperature~\cite{ogawa} defined as the mean square fluctuation of the
velocity, that is $T_G=\langle |{\mathbf v}- \langle {\mathbf v}
\rangle |^2 \rangle$ even though the velocity distribution could not
be Gaussian.

In the case of zero external forcing~\cite{goldhirsch} (i.e. the
so-called free cooling), granular temperature decreases in time and
goes asymptotically to zero when all the particles finally stop. When
an energy input feeds the system, instead, the granular temperature
may reach a stationary value and several modeling for the stochastic
driving have been proposed employing a
constant~\cite{Williams,twan,nostri,vannoije,Cafiero} or a
random~\cite{bar_tri} restitution coefficient.

In numerical simulations or in experiments, $T_G$ is often measured
taking the average $<\cdot>$ on the whole system. However strong
fluctuations in the local granular temperature are usually
observed~\cite{nostri}. Even considering only the global granular
temperature, i.e. for small spatial inhomogeneities, it is not clear
to which extent $T_G$ can be considered the ``temperature'' of the
system.

There are of course several possible paths to face this problem.  One
interesting point of view is that of investigating the response
properties of an external thermometer coupled to a granular
gas~\cite{exartier-peliti}. This corresponds to study the
fluctuation-dissipation properties of the
system~\cite{kurchanpeliti}. For a system slightly perturbed from its
stationary equilibrium state linear response theory allows to relate
the response to the correlation functions through the
fluctuation-dissipation relations~\cite{kubo_book}.

In the simplest case, given a perturbing field $\alpha$, a
fluctuation-dissipation relation relates the response of an observable
$B$ at the time $t$, after an impulsive perturbation at time $0$, to
the correlation of the observable $B$ and the field $A$, {\em
conjugated} to $\alpha$, measured in the unperturbed system,

\begin{equation}
\frac{\partial \langle B(t)\rangle}{\partial \alpha}= - \frac 1T
\frac \partial{\partial t} \langle B(t) A(0)\rangle,
\label{fdt-0}
\end{equation}

\noindent where $T$ is the equilibrium temperature of the system.

Recently Green-Kubo expressions for a homogeneous cooling granular gas
have been obtained~\cite{vannoije_fdt,dufty,brey} and numerical
simulations have confirmed their validity~\cite{lutsko}: cooling
granulars lack an equilibrium state, therefore the Homogeneous Cooling
State (which is characterized by scaling properties) is used as
reference state to be perturbed, but in this case the Green-Kubo
relations must be changed in order to keep into account new terms
arising from the time dependence of the reference state and the
non-conservative character of collisions (see~\cite{brey}). However,
in the case of steady state granular gases, a rigorous derivation of
Green-Kubo relations has not yet been performed to our knowledge.

In this paper we perform numerical investigations in order to check
the validity of standard Kubo formulas~\cite{kubo_1957,kubo_book} to
steady state inelastic gases.  We perform in particular two different
sets of numerical experiments on heated granular gases.  We choose two
different conjugated pairs of variables constituted by the
autocorrelation of a given variable and the corresponding response to
a perturbation applied to the gas. Kubo's formula, i.e. the
proportionality between response and the autocorrelation in the
unperturbed system, is verified to hold using the granular temperature
as the correct proportionality factor, at least for not too strong
inelasticities ($r > 0.5$). For stronger inelasticities (smaller
values of $r$) Kubo's formula can be verified with a different
effective temperature. It should be noted how the adopted simulation
scheme (i.e. Direct Simulation Monte Carlo) represents the numerical
implementation of the Boltzmann equation (given for example
in~\cite{nostri}) which assumes Molecular Chaos
hypothesis and therefore neglects short range correlations. For this
reason, for very strong inelasticities the Boltzmann equation (and the
DSMC scheme) cannot be considered realistic and one should use
Molecular Dynamics simulations.

These results differ from the analogous results obtained for dense
granular assemblies, where ``slow'' degrees of freedom thermalize at
an effective temperature which is far higher than the external imposed
temperature~\cite{bkls,kurchan}. However the gas-like state of
granular matter has nothing to do with these systems, as their
stationary state is governed by a rapid decay of the fluctuations and
the granular temperature turns out to be the right choice in the
description of linear response to slight perturbations.

The outline of the paper is as follows.  In section II we define the
model as well as its simulation scheme and give a brief review of
known results about the peculiar features of its non-equilibrium
stationary state. Section III reviews Kubo's formula and its
extension to non-Hamiltonian perturbations. In section IV we describe
numerical experiments using a sinusoidal shear and measuring as
response the density current. In section V we perform a diffusion
vs. mobility experiment upon a tracer (i.e. perturbing a single
particle). Finally section VI is devoted to the conclusions.

\section{The model}

We simulate a gas of $N$ identical particles of unitary mass in a
two-dimensional box of side $L=\sqrt{N}$ with periodic boundary
conditions. The particle collisions are inelastic: the total momentum
is conserved, while the component of the relative velocity parallel to
the direction joining the center of the particles is reduced to a
fraction $r$ (with $0 \leq r \leq 1$) of its initial value, lowering
in this way the kinetic energy of the pair:

\begin{subequations}
\label{inelastic_collision}
\begin{align}
\mathbf{v}_1'
&=\mathbf{v}_1-\frac{1+r}{2}((\mathbf{v}_1-\mathbf{v}_2)\cdot
\hat{\mathbf{n}}) \hat{\mathbf{n}} \\ \mathbf{v}_2'
&=\mathbf{v}_2+\frac{(1+r)}{2}((\mathbf{v}_1-\mathbf{v}_2)\cdot
\hat{\mathbf{n}}) \hat{\mathbf{n}}
\end{align}
\end{subequations}

\noindent where $\mathbf{v}_1'$ and $\mathbf{v}_2'$ are the velocities
of the colliding particles {\em after} the collision.

In the interval between two subsequent collisions, the motion of each
particle $i$ is governed by the following Langevin
equation~\cite{nostri}:

\begin{subequations}
\label{langevin}
\begin{align}
\frac{d}{d t}{\mathbf v}_i(t) & =  -\frac{{\mathbf v}_i(t)}{\tau_B}
+ \sqrt{\frac{2T_B}{\tau_B}}\mbox{\boldmath $\eta$}_i(t)  \label{langevin1} \\
\frac{d}{d t}{\mathbf x}_i(t) & =  {\mathbf v}_i(t)  \label{langevin2}
\end{align}
\end{subequations}

\noindent where the function $\mbox{\boldmath $\eta$}_i(t)$ is a
stochastic process with average $\langle \mbox{\boldmath $\eta$}_i(t) \rangle=0$ and
correlations
$\langle\eta^\alpha_i(t)\eta^\beta_j(t')\rangle=\delta(t-t')\delta_{ij}\delta_{\alpha
\beta}$ ($\alpha$ and $\beta$ being component indexes), i.e. a
standard white noise.  This means that each particle feels a hot fluid
with a temperature $T_B$ and a viscosity characterized by a time
$\tau_B$. 

The question about the most proper way of modeling a stochastic
driving is still open. Many authors for instance use
Eq.~\eqref{langevin} without viscosity~\cite{Williams}. This is
equivalent to the limit $\tau_B \to \infty$ and $T_B \to \infty$ with
keeping $D=T_B/\tau_B$ constant. In this limit long and short-range
correlations in the velocity and density fields have been
observed~\cite{vannoije}.  We have measured correlations in the
velocity field in the model with viscosity, concluding that they are
highly reduced by the viscous term that breaks Galilean invariance
(the frame $\mathbf{v}=0$ is preferred)~\cite{tesi}. Our choice of the
stochastic driving with viscosity has the following advantages: a) it
guarantees that in the elastic limit ($r=1$) the system, after a
transient time of the order of $\tau_B$, still reaches a stationary
state, characterized by a uniform density and a Gaussian distribution
of velocities whith temperature $T_B$; b) it is a heat bath with a
well defined finite temperature $T_B$ that can be compared with the
granular temperature $T_G$ and the effective temperature $T_{eff}$
measured by means of Fluctuation-Dissipation relations (see ahead).

\begin{figure}[ht]
\centerline {\includegraphics[clip=true,width=8cm,
keepaspectratio]{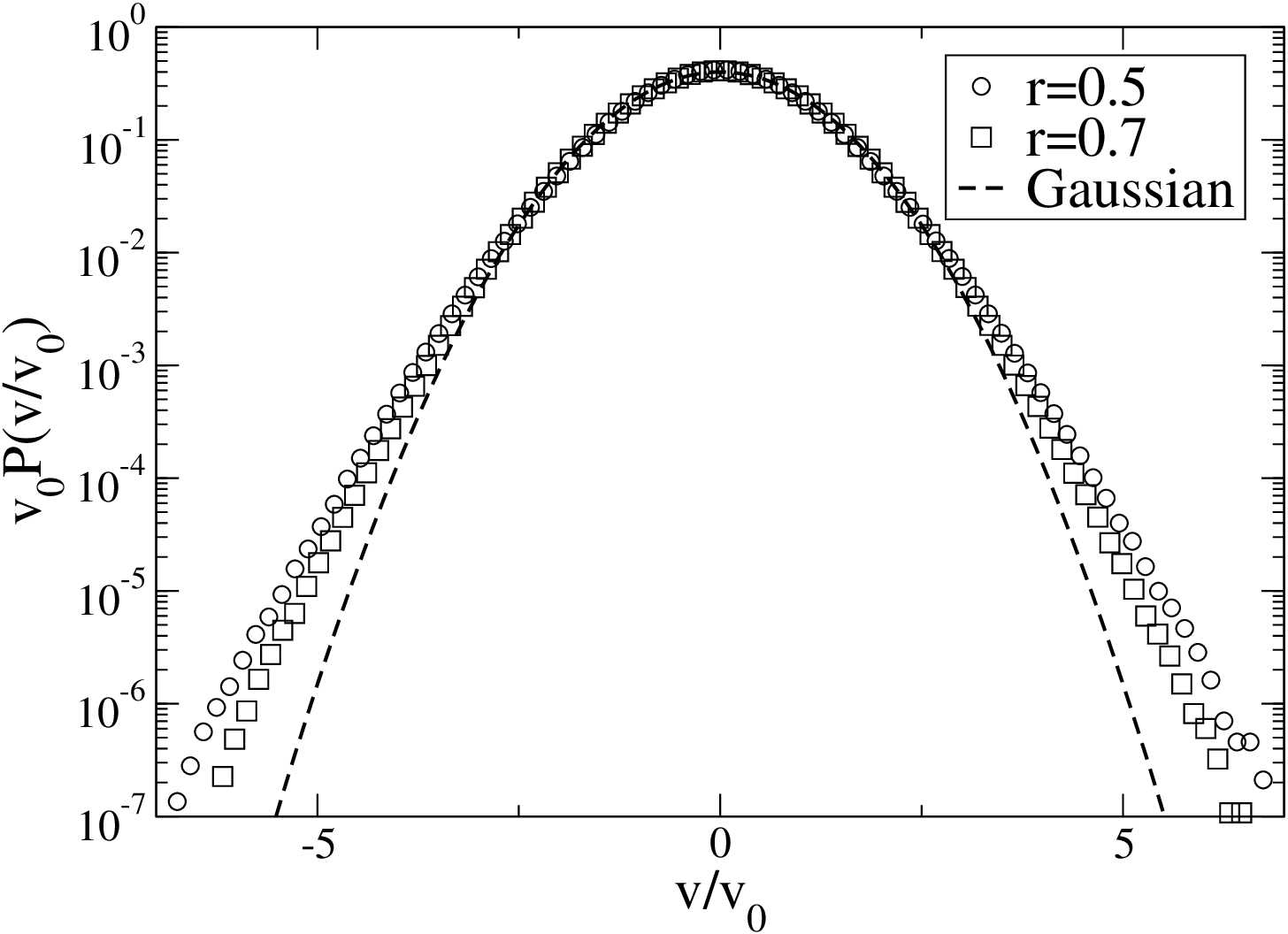}}
\caption{\label{dist_vel} 
Rescaled (in order to have unitary variance) distribution of the
horizontal component of the velocity of the gas $P(v_x)$ versus $v_x$,
with $N=500$, $L^2=N$, heat bath with $T_b=1$, $\tau_b=10$,
$\tau_c=1$, and different restitution coefficients. The Gaussian is
plot as a reference for the eye.}
\end{figure}

In spite of the drastic reduction of velocity correlations, this model
exhibits several interesting features which are parametrized by the
restitution coefficient $r$ and the ratio, $\tau_C/\tau_B$, between
the mean free time (the average interval of time between two
subsequent collisions of the same particle) and the viscosity time.
When $\tau_B > \tau_C$ and $r<1$ the system is in a non-equilibrium
stationary state with a granular temperature $T_G<T_B$. This state is
characterized~\cite{nostri} by fractal density clustering and
non-Gaussian distributions of velocities (see fig.(\ref{dist_vel}))
with a dependence upon $r$ and $\tau_C/\tau_B$.  This situation
persists when Direct Simulation Monte Carlo (DSMC)~\cite{bird} is used
to perform more rapid and larger simulations of the system. The DSMC
simulation scheme consists of a discrete time integration of the
motion of the particles. At each time step of length $\Delta t$ the
following operations are performed:

\begin{enumerate}

\item \label{free} {\em Free streaming}: Eq. \eqref{langevin} is
integrated for a time step $\Delta t$ disregarding possible
interactions among the particles.

\item
{\em Collisions}: every particle has a probability $p=\Delta t/\tau_c$
of undergoing a collision. Its collision mate is chosen among the
particles in a circle of fixed radius $r_B$ with a probability
proportional to its relative velocity. The unitary vector $\mathbf{n}$, which
should be parallel to the line joining the centers of
particles, is chosen randomly instead and the collision is performed using
rule~\eqref{inelastic_collision}.

\end{enumerate}


\section{Short review about Kubo's formula}

In this section we review the Kubo's formula, which must hold in the
case of equilibrium dynamics. First we show the case of
an Hamiltonian dynamics, and then for non-Hamiltonian equilibrium
dynamics of a system subjected to an (impulsive) shear force.
The first case is presented just for completeness and to set the
notations, the formulas presented in the second case will be
numerically checked for an elastic system and extended to the
inelastic case in the next sections.

\subsection{Kubo's formula for Hamiltonian systems}

For an Hamiltonian system, the temporal evolution of the phase space
distribution $f(p,q,t)$ is ruled by the Liouville equation:

\begin{equation} \label{liouville}
\frac{\partial}{\partial t}
f(p,q,t)=i[\mathcal{L}+\mathcal{L}_{ext}(t)]f(p,q,t)
\end{equation}
where $\mathcal{L}$ and $\mathcal{L}_{ext}$ are the Liouville
operators relative to the unperturbed Hamiltonian and to its
perturbation respectively. They are defined by means of classical
Poisson bracket:

\begin{equation}
i\mathcal{L}f=(\mathcal{H},f)=\sum \left( \frac{\partial
\mathcal{H}}{\partial q}\frac{\partial}{\partial p}-\frac{\partial
\mathcal{H}}{\partial p}\frac{\partial}{\partial q} \right) f
\end{equation}

The perturbation is assumed to be given by coupling a force
$\alpha(t)$ with an observable of the system $\hat{A}$, i.e.:

\begin{equation} \label{perturbation}
\mathcal{H}_{ext}=-\hat{A}(p,q) \alpha(t)
\end{equation}

The phase space distribution is assumed to be canonical (i.e. at
equilibrium) in the infinite past: $f(p,q,-\infty)=f_{eq}(p,q)$. An
approximate solution of~\eqref{liouville} to the first order in the
perturbation $i\mathcal{L}_{ext}$ is given by:

\begin{equation}
f(p,q,t)=f_{eq}(p,q)+\int_{- \infty}^t dt' \exp
[i(t-t')\mathcal{L}](\mathcal{H}_{ext}(t'),f_{eq}(p,q))+ \dotsc
\end{equation}

With this approximation, the deviation, due to the
perturbation~\eqref{perturbation}, of the expectation of a physical
quantity $\hat{B}$ can be written as the following convolution:

\begin{equation} \label{deviation}
\delta \hat{B}(t) = \langle \hat{B} \rangle_t - \langle \hat{B}
\rangle_{-\infty} = \int_{-\infty}^t dt' \Phi_{BA}(t-t')\alpha(t)
\end{equation}
where $\langle \dotsc \rangle_t$ denotes averages taken over the
ensemble given by $f(p,q,t)$; $\Phi_{BA}(t-t')$, called {\bf response
function}, represents the response $\delta \hat{B}(t)$ to the pulsed
force $\alpha(t)=\delta(t)$, and reads:

\begin{equation}
\Phi_{BA}(t)=\iint dp dq f_{eq}(p,q)(\Delta \hat{A}(p,q),\Delta
\hat{B}(p_t,q_t))=\langle (\Delta \hat{A}, \Delta \hat{B}(t)) \rangle
\end{equation}
where $\Delta {A}=\hat{A}-\langle A \rangle$ (and identically for
$\Delta \hat{B}$), while $(p_t,q_t)$ is the image of the initial phase
point $(p,q)$ determined by the total Hamiltonian (including the
perturbation).

Kubo has shown that the response function can be written in a simpler
form~\cite{kubo_1957,kubo_book}:

\begin{equation} \label{kubo_formula}
\Phi_{BA}(t)=\beta \langle \frac{\partial \Delta \hat{A}}{\partial
t}(0) \Delta \hat{B}(t) \rangle=-\beta \langle \Delta A(0)
\frac{\partial \Delta \hat{B}}{\partial t}(t) \rangle
\end{equation}

\noindent where $\beta$ is the inverse of the temperature.


\subsection{Fluctuation-Dissipation for non-Hamiltonian equilibrium systems 
(the case of shear force)}

\label{shear_section}

Let consider a gas of particles and define a non-conservative
perturbation (force) acting on particle $i$ placed at
$\mathbf{r}_i(t)$ at time $t$ as

\begin{equation} \label{forcing}
\mathbf{F}(\mathbf{r}_i,t)=\gamma_i \boldsymbol{\xi}(\mathbf{r}_i,t)
\;\;\;\;\; \mbox{with the properties} \;\;\; \boldsymbol{\nabla} \times \boldsymbol{\xi} \neq 0
\;\;\;  \boldsymbol{\nabla} \cdot \boldsymbol{\xi} = 0.
\end{equation}

\noindent For {\em small enough perturbation } and for any variable
(``observable'') $O(\mathbf{r})$ such that $\langle O
\rangle_{-\infty}=0$, equation~\eqref{deviation} with the Kubo
formula~\eqref{kubo_formula} reads~\cite{ciccotti,jackson}:

\begin{subequations}
\label{green-kubo}
\begin{align}
\langle        O(\mathbf{r})       \rangle_t        &=\beta       \int
d\mathbf{r}'\int_{-\infty}^t   dt'   \langle  O(\mathbf{r},t)   \sum_i
\gamma_i \dot{\mathbf{r}}_i(t') \delta\{\mathbf{r}'-\mathbf{r}_i(t')\}
\rangle_{-\infty}   \cdot  \boldsymbol{\xi}(\mathbf{r}',t')\\  \langle
\hat{O}(\mathbf{k})  \rangle_t  &=\beta  \int_{-\infty}^t dt'  \langle
\hat{O}(\mathbf{k},t)     \sum_i    \gamma_i    \dot{\mathbf{r}}_i(t')
\exp\{i\mathbf{k}\cdot   \mathbf{r}_i(t')\}   \rangle_{-\infty}  \cdot
\hat{\boldsymbol{\xi}}(\mathbf{k},t') \label{gk_fourier}
\end{align}
\end{subequations}
\noindent where the Fourier transform $G \to \hat{G}$ is defined as

\begin{equation}
\hat{G}(\mathbf{k})=\frac{1}{V}\int d\mathbf{r} e^{-i \mathbf{k} \cdot
\mathbf{r}} G(\mathbf{r})
\end{equation}

A force satisfying the properties (\eqref{forcing}) is for instance
given by:

\begin{equation}\label{force}
\boldsymbol{\xi}_{\bar{k}}(\mathbf{r},t)=
\begin{pmatrix}
0 \\ \Xi \exp(i \bar{k}_x x) \delta(t) 
\end{pmatrix}
\end{equation}
\noindent whose spatial Fourier transform reads:

\begin{equation}
\boldsymbol{\xi}_{\bar{k}}(\mathbf{k},t)=
\begin{pmatrix}
0 \\ \frac{\Xi}{V} \delta(\mathbf{k}-\bar{\mathbf{k}}) \delta(t)
\end{pmatrix}
\end{equation}
\noindent where $\bar{\mathbf{k}}=(\bar{k}_x,0)$ (having chosen
$k_x$ compatible with the periodic boundary conditions,
i.e. $k_x=2 n_k \pi/L_x$ with $n_k$ integer and $L_x$ the linear horizontal
dimension of the box where the particles move). Note that $\Xi$ must
have the dimensions of a momentum, i.e. of a velocity (taking unitary
masses). With this choice, eqs. \eqref{gk_fourier} becomes:

\begin{equation} 
\label{gk_shear}
\langle \hat{O}(\mathbf{k}) \rangle_t =\frac{\beta \Xi}{V} \langle
\hat{O}(\mathbf{k},t) \sum_i \gamma_i \dot{y}_i(0)
\exp\{i\mathbf{k}\cdot \mathbf{r}_i(0)\} \rangle_{-\infty}
\delta(\mathbf{k}-\bar{\mathbf{k}})
\end{equation}

\noindent If we now define the staggered $y$-current as:

\begin{subequations}
\begin{align}
J_y^{st}(\mathbf{r},t)& =  \sum_i \gamma_i \dot{y}_i(t)
\delta(\mathbf{r}-\mathbf{r}_i(t)) \\
\hat{J}_y^{st}(\mathbf{k},t)& =  \frac{1}{V}\sum_i \gamma_i \dot{y}_i(t)
\exp(-i \mathbf{k} \cdot \mathbf{r}_i(t)),
\end{align}
\end{subequations}
then, using this current as observable $O$ the relation \eqref{gk_shear} is written as:

\begin{equation} \label{gk_shear2}
\begin{split}
\langle \hat{J}_y^{st} (\mathbf{k},t) \rangle_t &=\frac{\beta
\Xi}{V^2} \langle \sum_{ij} \gamma_i \dot{y}_i(t) \exp(-i \mathbf{k}
\cdot \mathbf{r}_i(t)) \gamma_j \dot{y}_j(0) \exp\{i\bar{k}_x x_j(0)\}
\rangle_{-\infty} \delta(\mathbf{k}-\bar{\mathbf{k}}) \\ &=\beta \Xi
\langle \hat{J}_y^{st}(\mathbf{k},t)
\hat{J}_y^{st}(-\mathbf{k},0)\rangle_{-\infty}
\delta(\mathbf{k}-\bar{\mathbf{k}})
\end{split}
\end{equation}

A real linear combination of forces of the kind in eq.~\eqref{force}
is

\begin{equation}\label{real_force}
\boldsymbol{\xi}(\mathbf{r},t)=\frac{1}{2}(\boldsymbol{\xi}_{\bar{k}}
(\mathbf{r},t)+\boldsymbol{\xi}_{-\bar{k}}(\mathbf{r},t))=
\begin{pmatrix}
0 \\ \Xi \cos (\bar{k}_x x) \delta(t) 
\end{pmatrix}
\end{equation}

With this choice of the perturbation, the relation \eqref{gk_shear2}
becomes:

\begin{equation} \label{gk_shear3}
\langle \hat{J}_y^{st} (\mathbf{k},t) \rangle_t =\frac{\beta \Xi}{2}
\langle \hat{J}_y^{st}(\mathbf{k},t)
\hat{J}_y^{st}(-\mathbf{k},0)\rangle_{-\infty}
(\delta(\mathbf{k}-\bar{\mathbf{k}})+\delta(\mathbf{k}+\bar{\mathbf{k}})).
\end{equation}

\noindent This is a {\em fluctuation-dissipation} relation which
expresses the fact that the response of the
$\bar{\mathbf{k}}$-component of the transverse current to the
perturbing field in eq.~\eqref{real_force} is proportional to the
auto-correlation of the same transverse current measured in the system
{\em without perturbation}.

The real part of the response calculated at
$\bar{\mathbf{k}}$ (per unit of perturbing field) is directly
computable and reads:

\begin{equation} 
\label{fd1}
\begin{split}
Re \left[ \frac{V}{\Xi} \langle \hat{J}_y^{st} (\mathbf{\bar{k}},t)
\rangle_t \right] & = \frac{1}{\Xi}\langle \sum_i \gamma_i
\dot{y}_i(t) \cos(\bar{k}_x x_i(t)) \rangle_t.
\end{split}
\end{equation}

From eq.(\ref{gk_shear3}) one obtains the relation:

\begin{equation} 
\label{fd2}
\begin{split}
\frac{1}{\Xi}\langle \sum_i \gamma_i \dot{y}_i(t) \cos(\bar{k}_x
x_i(t)) \rangle_t =\frac{\beta}{2} &\langle \sum_{ij} \gamma_i
\gamma_j \dot{y}_i(t) \dot{y}_j(0) \cos \{\bar{k}_x[x_i(t)-x_j(0)] \}
\rangle_{-\infty}.
\end{split}
\end{equation}

\section{Fluctuation-Dissipation measure I: shear vs. current}

\begin{figure}[ht]
\centerline {\includegraphics[clip=true,width=8cm,
keepaspectratio]{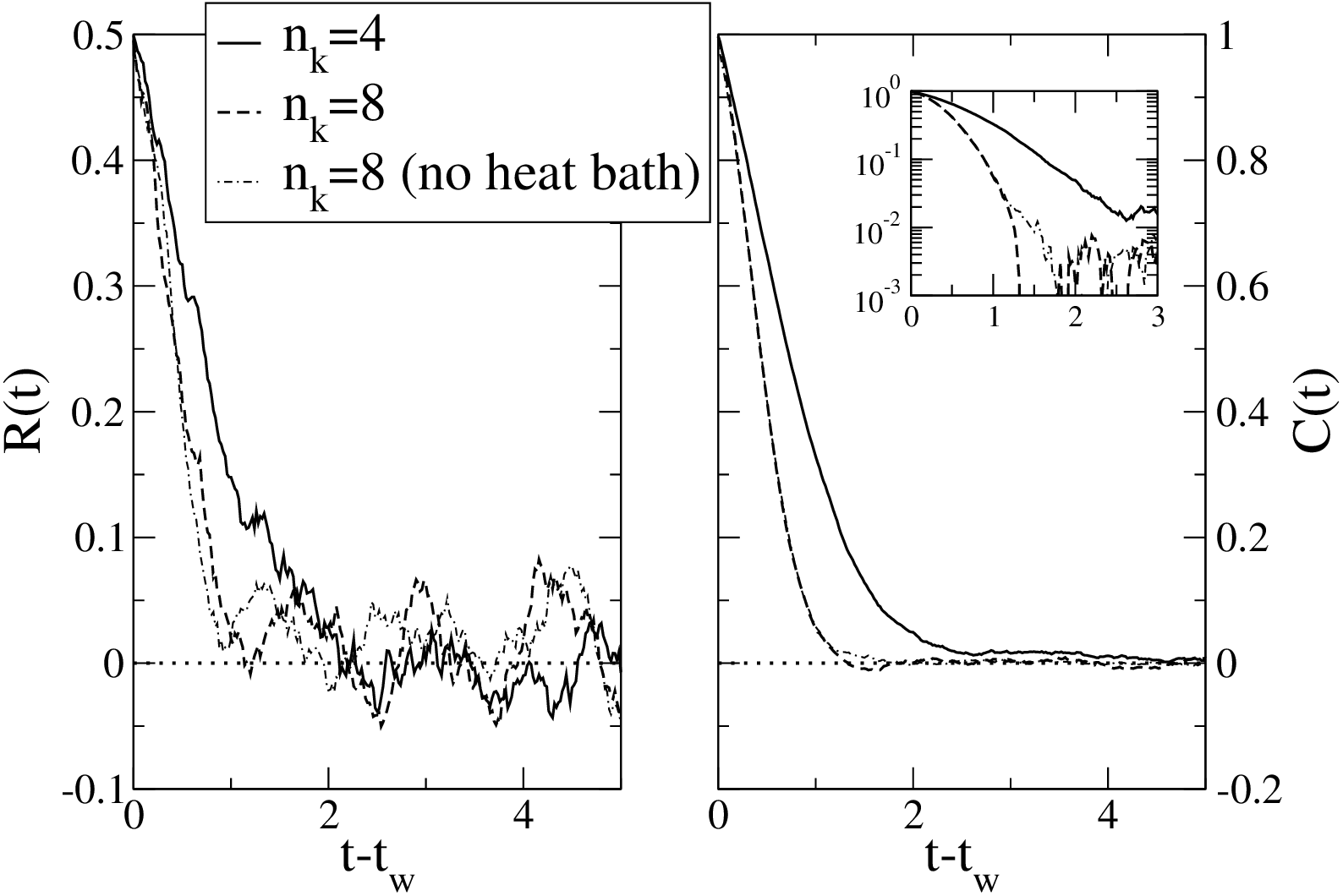}}
\caption{\label{cicc_rc} Left: time dependent response to the
impulsive shear perturbation defined in~\ref{forcing}
and~\ref{real_force}: $R(t-t_w)$ vs. $t-t_w$ for three simulations
with elastic systems, one without thermal bath and two with thermal
bath, and with different choices of the wave number $n_k$ of the
perturbation. Right: time correlation function $C(t-t_w)$ vs. $t-t_w$
for the same systems. In the simulations $r=1$, $N=500$, $\tau_c=1$,
and for the two cases with the heat bath $T_b=1$ and $\tau_b=10$. The
applied force has $\Xi=0.01$, $\tau_w=100$.  The averages have been
obtained over $10000$ realizations.}

\centerline {\includegraphics[clip=true,width=8cm,
keepaspectratio]{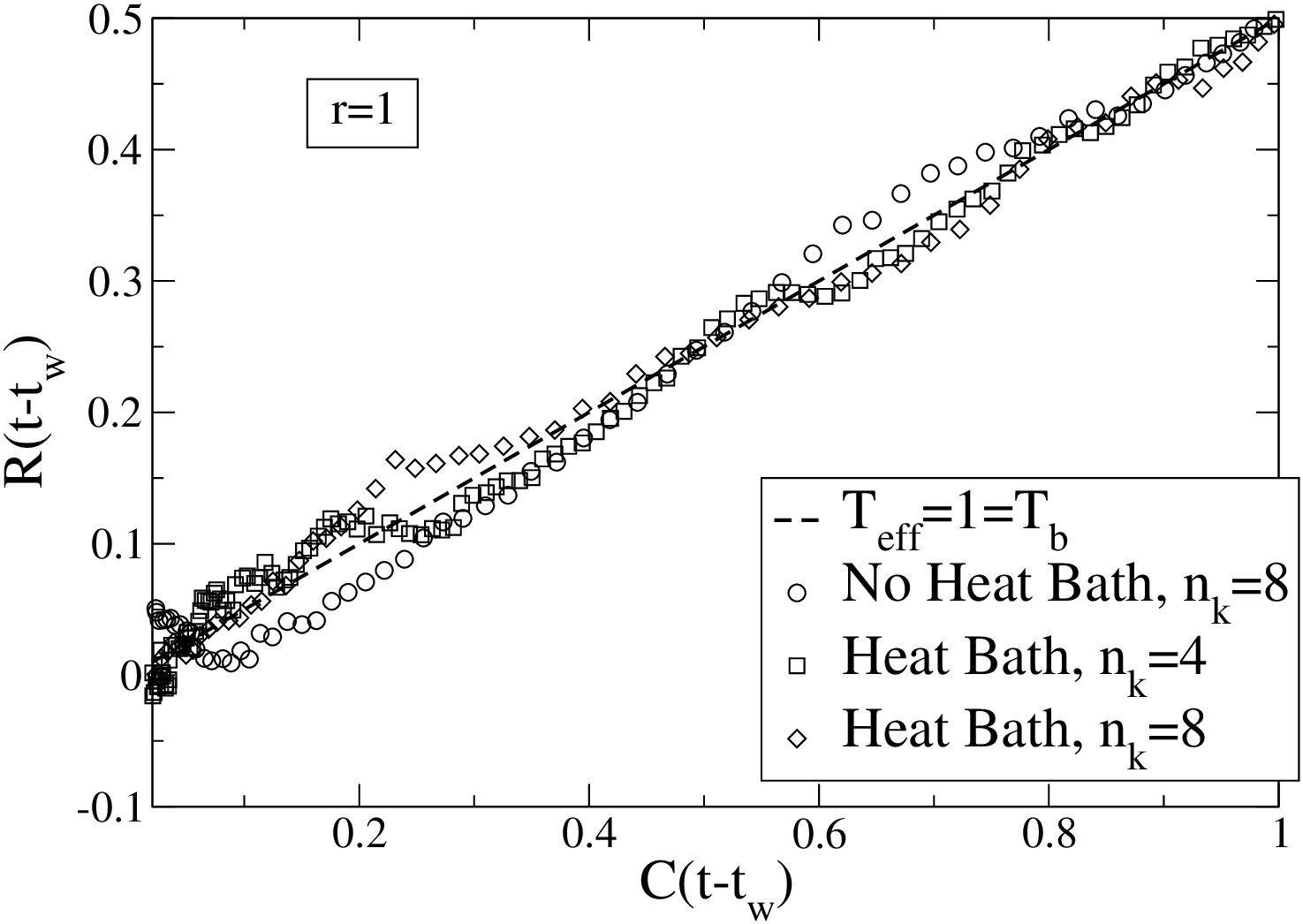}}
\caption{\label{cicc_el} Parametric plot of $R(t-t_w)$ vs. $C(t-t_w)$
for the numerical experiment of type $I$ (impulsive shear
perturbation) with $r=1$, with or without heating bath, and for
different choices of the wave number $n_k$ of the perturbation. The
initial temperature for the case without heat bath is chosen to be $1$
while $T_b=1$ ad $\tau_b=10$ for the two cases with the heat bath.
$N=500$, $\tau_c=1$, $\Xi=0.01$, $n_k=8$, with average over $10000$
realizations, using $t_w=100$.}
\end{figure}

The first set of measures we have performed has been aimed to verify
relation~\eqref{fd2} for a system as described in the previous
paragraph. In order to do this we have used the following recipe
proposed in~\cite{ciccotti}:

\begin{enumerate}

\item
Initialize system $U$ with random positions $\{\mathbf{r}_i^U(0)\}$
and random velocities $\{\dot{\mathbf{r}}_i^U(0)\}$.

\item
Let it evolve with the unperturbed dynamics until time $t_w$ which
must be chosen larger than the largest characteristic time of the
system (e.g. $\tau_c$ or $\tau_b$). The unperturbed dynamics consists
of the time-discretized ($\Delta t$) integration of the Langevin
equation~\eqref{langevin}:

\begin{equation} \label{unperturbed}
\mathbf{v}_i(t+\Delta t) = \mathbf{v}_i(t) - \frac{\Delta t}{\tau_b}
\mathbf{v}_i(t) + \sqrt{\frac{2T_b \Delta t}{\tau_b}} \mathbf{R}(t)
\end{equation}
\noindent plus inelastic collisions with parameter $r$ (restitution
coefficient). The collisional step is separated from the Langevin step
and is implemented by means of a local Monte Carlo, i.e. random choice
of pairs to collide inside a region of diameter $r_{B}$; the collision
probability is proportional to a fixed collision frequency $1/\tau_c$
and to the relative velocity of the particles; $\tau_c$ is chosen to
be compatible with an homogeneous gas-like dynamics, i.e. $\tau_c
\approx r_{B}/\sqrt{\langle v^2 \rangle_{-\infty}}$.

\item 
At time $t_w$ a copy of system $U$ is created (and named $P$) and the
vectors $\{\dot{y}_i(t_w)\}$ and $\{x_i(t_w)\}$ memorized in order to
be used in the computation of the auto-correlation.

\item \label{perturbation_step} The system $U$ is let evolve with the
unperturbed dynamics. The system $P$ is made evolve with the
additional forcing described in eqs. \eqref{forcing} and
\eqref{real_force} for only the time step $[t_w,t_w+\Delta t]$,
i.e. the equation for the update of velocities in this particular step
being:

\begin{equation} \label{perturbed_force}
\mathbf{v}_i^P(t_w+\Delta t) = \mathbf{v}_i^P(t_w) - \frac{\Delta
t}{\tau_b} \mathbf{v}_i^P(t_w) + \sqrt{\frac{2T_b \Delta t}{\tau_b}}
\mathbf{R}(t_w) + \gamma_i
\begin{pmatrix}
0 \\ \Xi \cos [\bar{k}_x x_i^P(t_w)]
\end{pmatrix}
\end{equation}
\noindent with $\bar{k}_x=2\pi n_k/L_x$. Note again that the
perturbation intensity $\Xi$ has exactly the dimensions of a velocity.

\item
The dynamics of systems $U$ and $P$ are thereafter followed in the
unperturbed style, i.e. using eq. \eqref{unperturbed}. The functions
to be measured are:

\begin{subequations}
\begin{align}
R(\tau) &= \frac{1}{N \Xi} \sum_i \gamma_i \dot{y}_i^P(t_w+\tau)
\cos(\bar{k}_x x_i^P(t_w+\tau)) \\ C(\tau) &= \frac{1}{N} \sum_{ij}
\gamma_i \gamma_j \dot{y}_i^U(t_w+\tau) \dot{y}_j^U(t_w) \cos
\{\bar{k}_x[x_i^U(t_w+\tau)-x_j^U(t_w)] \},
\end{align}
\end{subequations}
\noindent where $\tau = t -t_w$.  It is expected that $R(0)=1/2$ and
$C(0)=\langle v^2 \rangle$, while $R(\infty)=C(\infty) \to 0$.

\item
The above steps are repeated for many different realizations (or even
in the same realization, provided its length is much longer than the
typical correlation time) and the averages over those realizations of
$R(t-t_w)$ and $C(t-t_w)$ are computed (see
figure~\ref{cicc_rc}). 

\end{enumerate}

To the aim of checking the whole numerical machinery, we first
consider the $r=1$ elastic case.
In all the cases investigated we have checked the linearity of the
response by changing the perturbation amplitude in the range $\Xi \in
[0.005,0.05]$.
Within the specific framework chosen for the
observables, the Kubo formula to be verified is given by:

\begin{equation} \label{viscosity}
\langle R(t-t_w) \rangle=\frac{\beta}{2} \langle C(t-t_w) \rangle
\end{equation}

We have performed the following experiments:

\begin{itemize}

\item
gas with elastic interactions and absence of thermal bath (see
figure~\ref{cicc_el}).

\item
gas with elastic interactions {\em with} the thermal bath (see
figure~\ref{cicc_el}).

\end{itemize}

From figure~\ref{cicc_rc} it can already be appreciated that response
and autocorrelations in the elastic gas decay on a time of the order
of $\tau_c$ (however this decay is not exponential, as can be observed
in the inset of the figure). The parametric plot of the two curves in
figure~\ref{cicc_el} shows the perfect agreement with
equation~\eqref{viscosity} using $\beta=1/T_b$. In this case, of
course, $T_b \equiv T_G$.

We have repeated the same measures on the
gas with restitution coefficient $r<1$, i.e.

\begin{figure}[ht]
\centerline {\includegraphics[clip=true,width=8cm,
keepaspectratio]{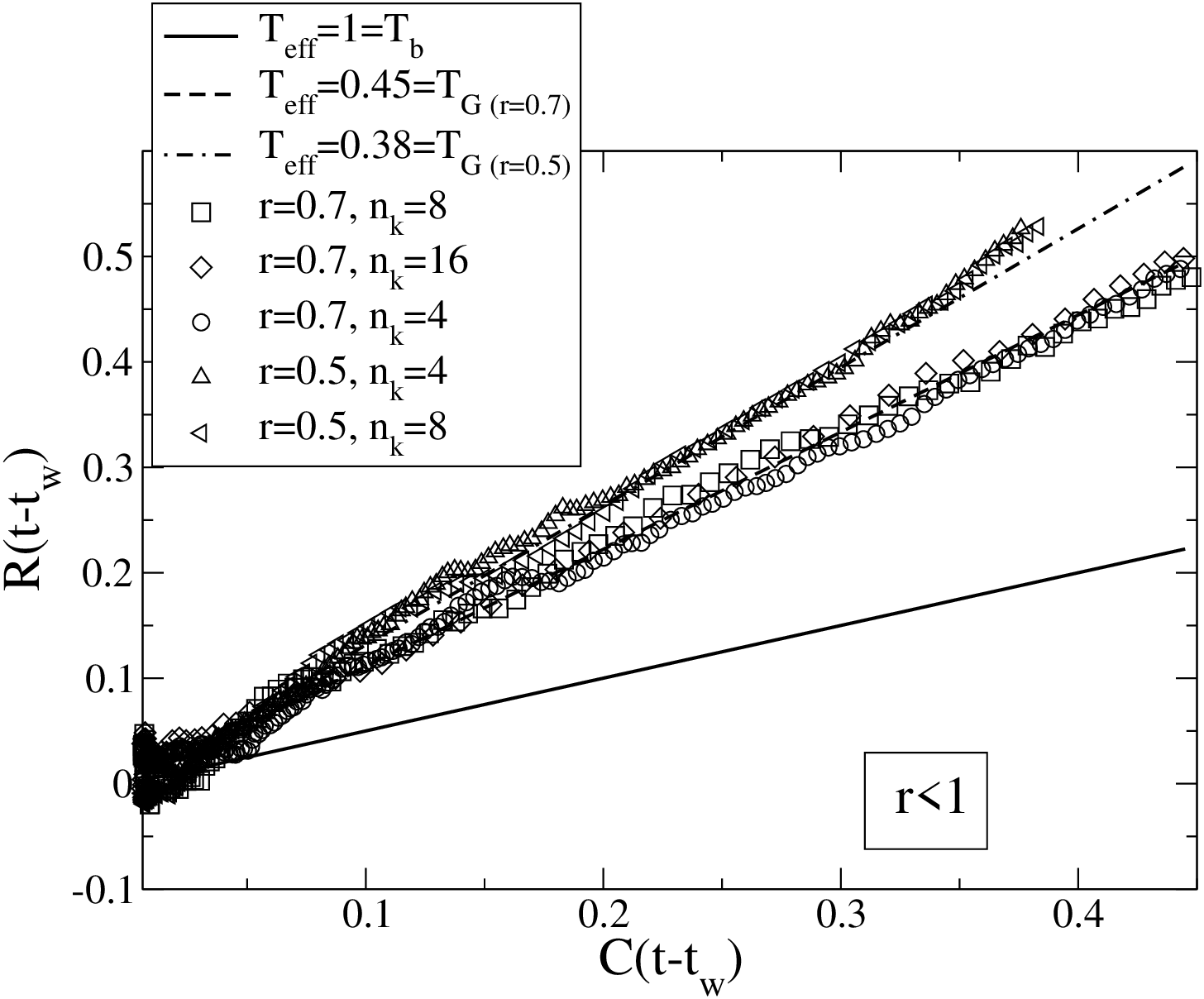}}
\caption{\label{cicc_inel} Parametric plot of $R(t-t_w)$
vs. $C(t-t_w)$ for the numerical experiment of type $I$ (impulsive
shear perturbation) with $r<1$, with heating bath, and for different
choices of the wave number $n_k$ of the perturbation. $T_b=1$ and
$\tau_b=10$, $N=500$, $\tau_c=1$, $\Xi=0.01$, $n_k=8$, with averages
over $10000$ realizations, using $t_w=100$.}
\end{figure}

\begin{itemize}
\item
gas with inelastic interactions with the bath (in this case the bath
is essential, to avoid cooling), see figure~\ref{cicc_inel}.
\end{itemize}

obtaining again a very good agreement with equation~\eqref{viscosity} using
$\beta=1/T_G$. 
This is the main finding in this set of numerical
experiments: even if the gas is out of equilibrium, being driven by a
thermal bath at temperature $T_b$, its unperturbed autocorrelation is
still proportional to the linear response, and its effective temperature measured
by means of Fluctuation-Dissipation theory is exactly the granular
temperature $T_G$. This is true for every inelasticity $r \ge 0.5$. 

At lower inelasticities we have still obtained an agreement with
equation~\eqref{viscosity} but with a $T_{eff}=1/\beta \neq T_G$: we
do not think that these measures can be interesting, as the DSMC
algorithm is not reliable at such values of $r$. A real system of
inelastic particles with $r<0.5$ should exhibit, even at low
densities, a high degree of clusterization which cannot be observed
within the DSMC framework. 

It also must be noted that relation~\eqref{viscosity} is verified for
many values of the wave number $n_k$, i.e. the system does not show a
scale dependent effective temperature.

\section{Fluctuation-Dissipation measure II: diffusion}

Another independent confirmation of the validity of the modified Linear Response
theory for granular gas comes from the study of the diffusion
properties, i.e. of the large time behavior of the mean squared
displacement $B(t,t_w)= \langle |\mathbf{r}(t)-\mathbf{r}(t_w)|^2
\rangle \sim 2D(t-t_w)$. In this case the Einstein relation is
expected to hold $D=<|\mathbf{v}|^2>\tau_{corr}$, where $\tau_{corr}=
\beta \int d\tau \langle v(t_w+\tau) v(t_w) \rangle$: this relation however,
often addressed as a sort of FD relation, is always verified and just
represents a check of the correctness of the simulation. Instead some
surprise could arise from mobility measurements: a small static drag
force (switched on at time $t_w$) of intensity $\Xi$ (in the direction
of the unitary vector $\hat{x}$ of the $x$ axis) is applied to a given
particle (tracer, e.g. particle with index $0$ and position $\mathbf{r}_0$). The tracer reaches,
as a result of the viscous force generated by the gas surrounding it,
a limit constant velocity such that $\chi(t,t_w) \equiv \langle
|(\mathbf{r}_0(t)-\mathbf{r}_0(t_w))
\cdot \hat x| \rangle \sim \Xi \mu t$, where $\mu$ is the mobility
which is expected to be related to the diffusion coefficient through
the Einstein relation $\mu=D/\langle \mathbf{v}_x^2 \rangle = 2 D /T $
(if the force is applied on the direction $x$ in the two-dimensional
system).

\begin{figure}[ht]
\centerline {\includegraphics[clip=true,width=8cm,
keepaspectratio]{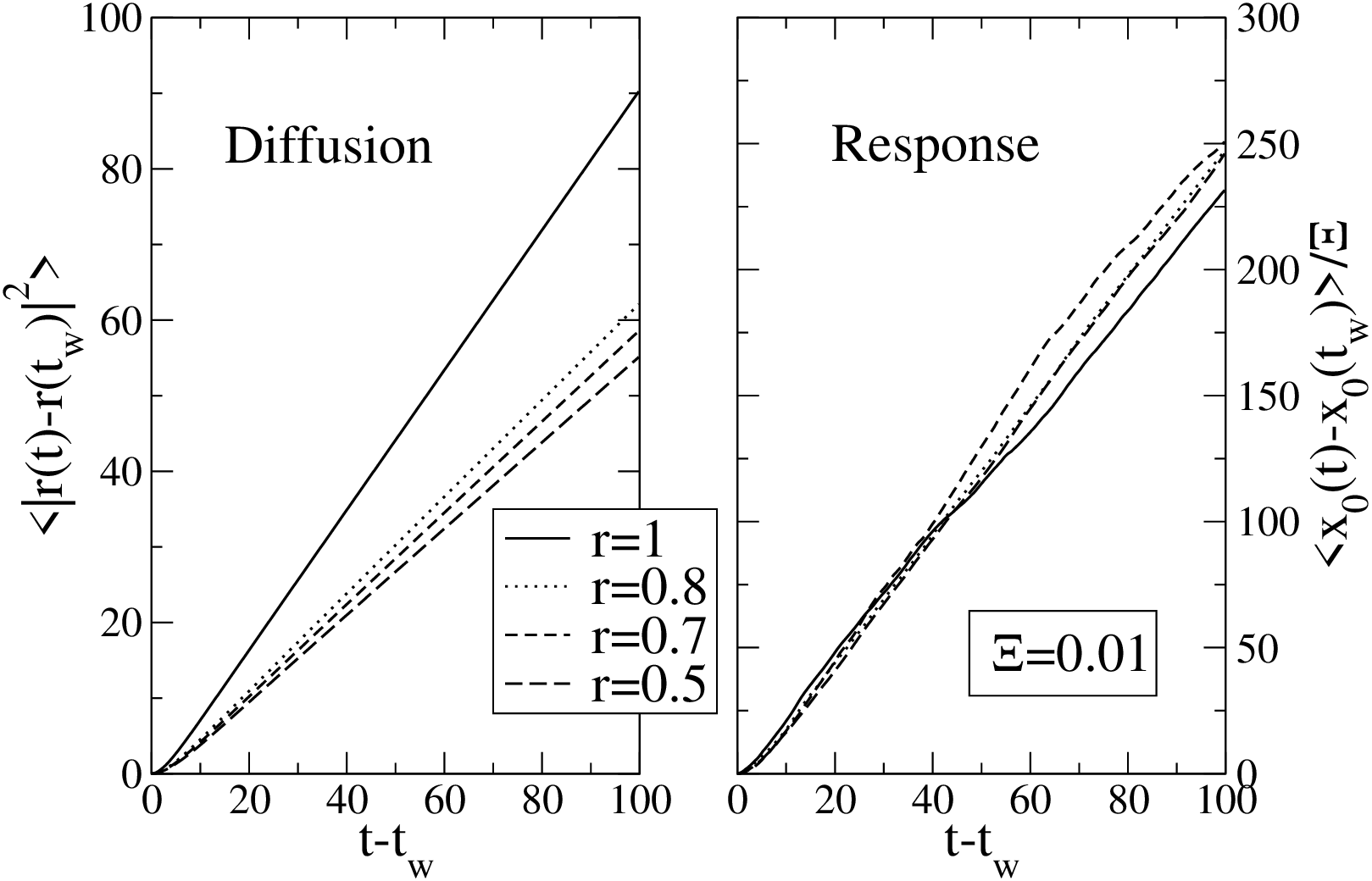}}
\caption{\label{diff_rc} Simulations of systems coupled to a thermal
bath with elastic or inelastic collisions. $N=500$, $\tau_c=1$,
$T_b=0.1$ and $\tau_b=10$, $\Xi=0.01$, $\tau_w=100$. The results are
obtained by averaging over $10000$ realizations. Left: mean squared
displacement $B(t,t_w)$ vs. $t-t_w$. Right: Integrated response
$\chi(t,t_w)$ to a constant force applied to the particle numbered as
$0$ vs. $t-t_w$.}
\end{figure}

In our experiments we have checked the linearity of the relation
between $\langle |x_0(t)-x_0(t_w)| \rangle\equiv \langle
|(\mathbf{r}_0(t)-\mathbf{r}_0(t_w)) \cdot \hat x|\rangle$ and
$\langle |\mathbf{r}(t)-\mathbf{r}(t_w)|^2 \rangle$ (see
fig.~\ref{diff_rc}). In particular if Kubo's formula was valid one
should have:

\begin{equation} \label{r_d}
\frac{ \langle |x_0(t)-x_0(t_w)| \rangle
}{\Xi} = \beta \frac{ \langle |\mathbf{r}_i(t)-\mathbf{r}_i(t_w)|^2 \rangle
}{4},
\end{equation}
with $\beta=1/T$ if the system is in thermodynamic equilibrium at
temperature $T$. 

In all the simulations we have checked the linearity of the response
by changing the perturbation amplitude in the range $\Xi \in [0.005,0.05]$.

Figure~\ref{diff_rc} shows the mean squared displacement (in the
unperturbed system) and the $x$ displacement of the tracer (when it is
accelerated) divided by the intensity of the perturbing force versus
time: it can be appreciated how both these quantities grow linearly
with time, defining the diffusion coefficient and the mobility. 

\begin{figure}[ht]
\centerline {\includegraphics[clip=true,width=8cm,
keepaspectratio]{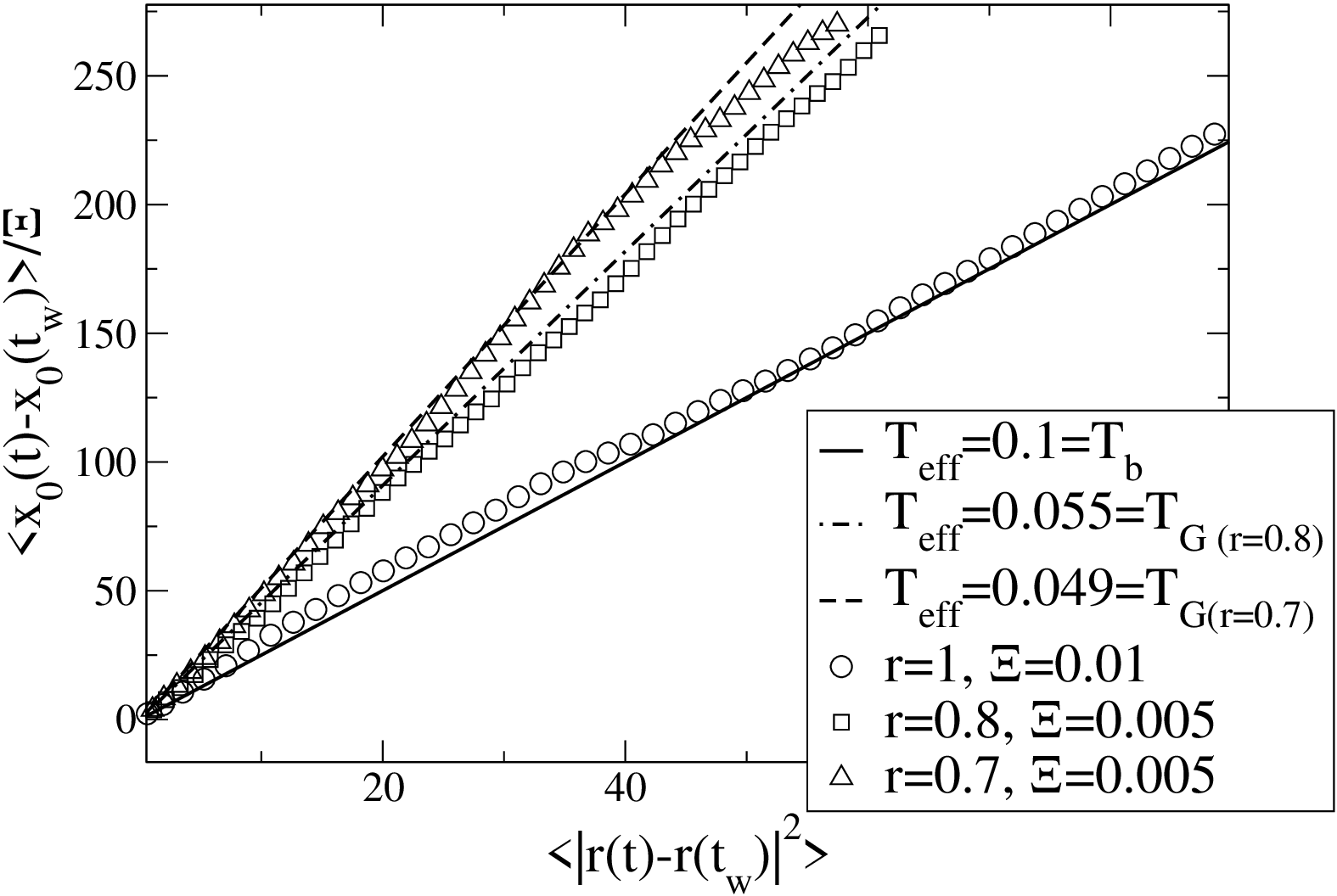}}
\caption{\label{diff_1} Parametric plot of $\chi(t,t_w)$
vs. $B(t,t_w)$ for the numerical experiment of type $II$ (constant
force applied on one particle) with $r=1$, $r=0.8$ and $r=0.7$, with heating
bath, and for different choices of the intensity $\Xi$ of the
perturbation, using $T_b=0.1$ ad $\tau_b=10$, $N=500$, $\tau_c=1$,
$t_w=100$. The results are obtained by averaging over $10000$
realizations.}
\end{figure}

\begin{figure}[ht]
\centerline {\includegraphics[clip=true,width=8cm,
keepaspectratio]{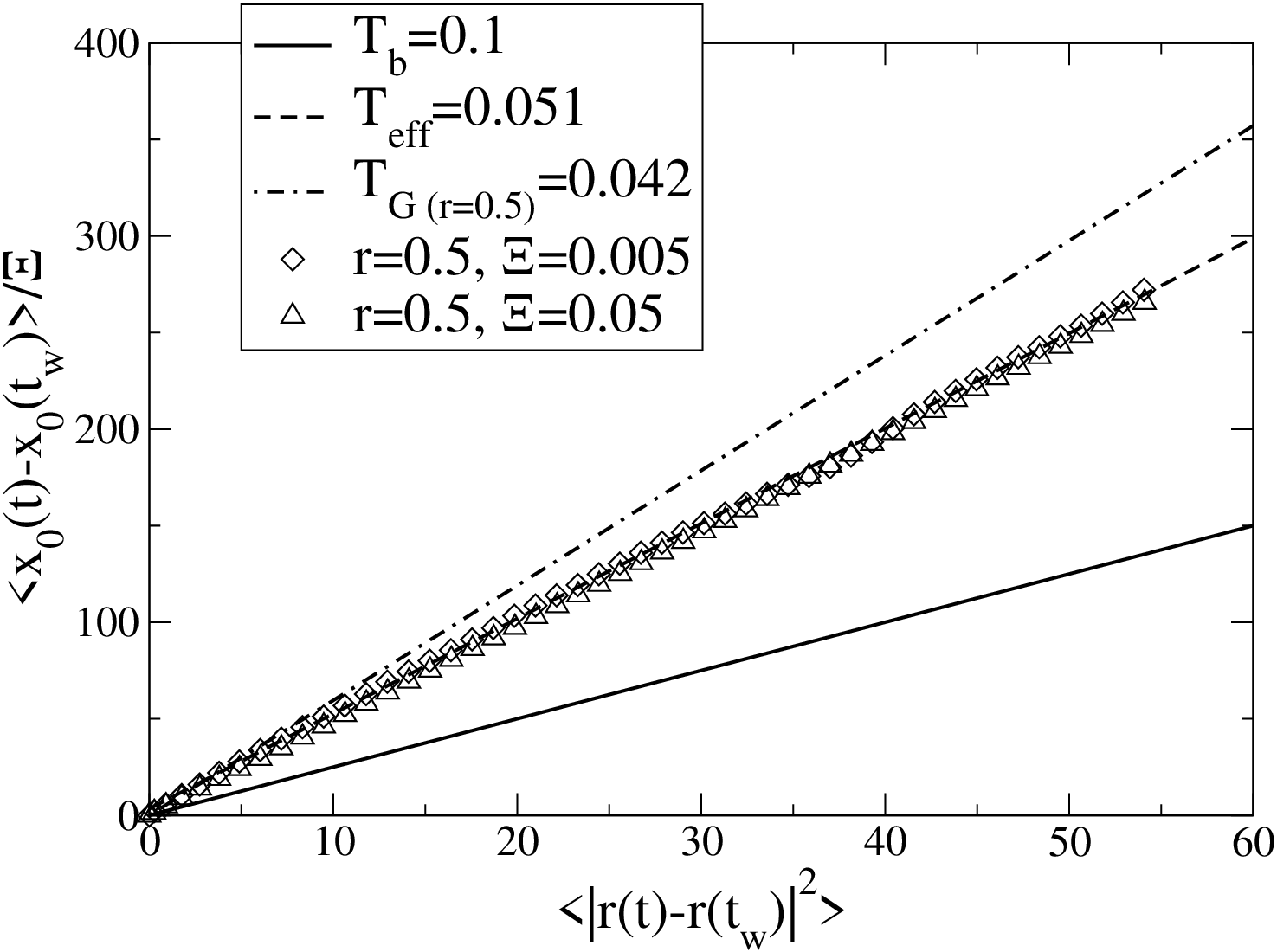}}
\caption{\label{diff_2} Parametric plot of $\chi(t,t_w)$
vs. $B(t,t_w)$ for the numerical experiment of type $II$ (constant
force applied on one particle) with $r=0.5$, with heating bath, and
for different choices of the intensity $\Xi$ of the perturbation,
using $T_b=0.1$ ad $\tau_b=10$, $N=500$, $\tau_c=1$, with average on
$10000$ realizations, using $t_w=100$. This case $r=0.5$ exhibits a
slope compatible with $T_{eff}=0.051$ which is above $T_G$.}
\end{figure}

Figures (\ref{diff_1},\ref{diff_2}) report the parametric plots of the
response to the force versus the mean squared displacement in the
unperturbed system, showing how relation~\eqref{r_d} is very well
satisfied at different inelasticities. What can be observed in this
kind of measures is a departure from the relation $\beta=1/T_G$
already at inelasticity $r=0.5$ (i.e. where the check of
fluctuation-dissipation theorem with the first type of measures still
was positive): it appears that $\beta=1/T_{eff}$ with
$T_b>T_{eff}>T_G$. This breakdown of fluctuation-dissipation relation
should be related to cluster formation (i.e. spatial lack of
homogeneity), which is present even in the case of DSMC solutions of
Boltzmann equation (see for example~\cite{nostri}). This violation is
more pronounced than in the previous experiment, therefore appearing
earlier, because this is a ``local'' measurement of FD, i.e. only one
particle is perturbed: this means that the fluctuations of the local
granular temperature strongly influence the trajectory of the tracer
and its statistical properties. 

\section{Conclusions}

In conclusion in this paper we have investigated the validity of
Kubo's relations in Driven Granular Gases in two dimensions.  We have
compared in particular two sets of measures. On the one hand we have
measured the proportionality factor between the response of the
staggered density current to an impulsive shear forcing and its
autocorrelation function in the unperturbed system. On the other hand
we have monitored the velocity of a tracer by checking the
proportionality between its response to a small constant force,
switched on at time $t_w$, and its mean squared displacement in the
unperturbed stationary state.  

In both cases a proportionality is observed, in analogy with the
Linear Response theory for equilibrium dynamics. Furthermore the
proportionality factor in the Kubo formulas is equal to the inverse of
th granular temperature (at least for the restitution coefficients
larger than $r>0.5$), which hence plays the role of the equilibrium
temperature in the elastic case.  It is important to remark how these
results are recovered by two completely independent measurement
schemes.

Several remarks are in order.  First of all, though a granular gas is
a non-trivial out-of-equilibrium system, from the point of view of its
thermodynamics it exhibits properties which seem much simpler than the
corresponding properties observed in a compacting granular medium.  In
this case in fact apparently no slow modes are present or at least
their presence does not give raise to the existence of several
effective temperatures depending on the time-scales
investigated~\cite{kurchanpeliti}. On the other hand the
proportionality factor between response and autocorrelation we have
found cannot be considered a temperature from the point of view of
equilibrium thermodynamics since it does not rely on any known
statistical ensemble. 

Another point to stress concerns the validity of the zeroth principle
of thermodynamics. The question that immediately arises could be
summarized as follows: if the granular temperature represents the
correct temperature from the point of view of the
Fluctuation-Dissipation Theorem, should we expect it rules the
thermalization properties of two different granular gases put in
contact? As already mentioned the answer to this question is far from
being trivial as witnessed by all recent results obtained on
mixtures~\cite{garzo,Huilin,losert,menon,Wildman,MontaneroShear,Clelland,
MontaneroHCS,puglisi2,barrat_trizac,martin} where a lack of
equipartition is observed. However these results (validity of
fluctuation-dissipation relations and lack of equipartition) can
coexist simply because heated granular gases have not
only a thermal source but also a thermal sink (dissipative collisions)
and therefore any zero principle should be stated in terms of a balance
equation among energy fluxes instead of a strict equivalence between
temperature of systems in contact. 

We again remark, also, that there are other common ways of driving a
granular gas in a stationary state, e.g. stochastic driving without
viscosity~\cite{vannoije}, stochastic restitution coefficient
models~\cite{bar_tri}, multiplicative noise models~\cite{Cafiero} and
so on. In some of these models a more pronounced departure from
homogeneity (for example correlations in the velocity
field~\cite{vannoije}) and therefore a breakdown of
fluctuation-dissipation relations should be investigated.

{\Large Acknowledgements} The authors are grateful to S. Roux a
preliminary and enlightening discussion, and to A. Barrat and
E. Trizac for many interesting discussions as well as a critical
reading of the manuscript. This work has been partially supported by
the European Network-Fractals under contract No. FMRXCT980183.
A. B. acknowledges support from the INFM Giovani Valenti
fellowship. V. L. and A. P. acknowledge support from the INFM {\em
Center for Statistical Mechanics and Complexity} (SMC).



\begin{thebibliography}{99}

\bibitem{reviews} H.M.~Jaeger, S.R.~Nagel and R.P.~Behringer, {\em
Rev. Mod. Phys.} {\bf 68}, 1259 (1996); Proceedings of the NATO
Advanced Study Institute on {\it Physics of Dry Granular Media}, Eds.
H. J. Herrmann {\it et al}, Kluwer Academic Publishers, Netherlands
(1998); A. Mehta and T.C. Halsey Eds., Proceedings of the conference
"Challenges in Granular Physics", ICTP Trieste August 2001, {\em
Advances in Complex Systems} Vol.4 (2001).

\bibitem{luding} T.~P\"oschel and  S.~Luding (Eds.),  {\em Granular Gases},
  Springer, Berlin (2001).

\bibitem{ogawa} Ogawa S., Proc. of US-Japan Symp. on {\em Continuum
Mechanics and Statistical Approaches to the Mechanics of Granular
Media}, eds. S.C. Cowin and M. Satake (Gakujutsu Bunken Fukyu-kai,
1978), p.208.

\bibitem{goldhirsch} I. Goldhirsch and G. Zanetti, {\em
Phys. Rev. Lett.}  {\bf 70}, 1619 (1993).

\bibitem{Williams} D.R. Williams and F.C. MacKintosh, {\em Phys. Rev.
E} {\bf 54}, R9 (1996).

\bibitem{twan} T.P.C. van Noije and M.H. Ernst, {\em Gran. Matter}
{\bf 1}, 57 (1998).

\bibitem{nostri} A. Puglisi, V. Loreto, U. M. B. Marconi, A. Petri,
and A. Vulpiani, {\em Phys. Rev. Lett.} {\bf 81}, 3848 (1998);
A. Puglisi, V. Loreto, U. M. B. Marconi, and A. Vulpiani, {\em
Phys. Rev. E} {\bf 59}, 5582 (1999).

\bibitem{vannoije} T. P. C. van Noije, M. H. Ernst, E. Trizac, and
I. Pagonabarraga, {\em Phys. Rev. E} {\bf 59}, 4326 (1999);
I. Pagonabarraga, E. Trizac, T. P. C. van Noije, {\em Phys. Rev. E}
{\bf 65}, 011303 (2002).

\bibitem{Cafiero} R. Cafiero, S. Luding and H.J. Herrmann, {\em
Phys. Rev. Lett.} {\bf 84}, 6014 (2000).

\bibitem{bar_tri} Alain Barrat, Emmanuel Trizac and Jean-No\"el Fuchs,
{\em Eur. Phys. J. E} {\bf 5}, 161 (2001).

\bibitem{exartier-peliti} R. Exartier and L. Peliti, {\em
Eur. Phys. J. B} {\bf 16}, 119-126 (2000).

\bibitem{kurchanpeliti} L.F. Cugliandolo, J. Kurchan and L. Peliti,
{\em Phys. Rev. E} {\bf 55}, 3898-3914 (1997).

\bibitem{kubo_book} R. Kubo, M. Toda and N. Hashitsume, {\em
Statistical Physics II}, Springer-Verlag, Berlin (1991).

\bibitem{dufty}
J. W. Dufty and V. Garz\'o, {\em J. Stat. Phys.} {\bf 105}, 723 (2001).

\bibitem{brey}
J. W. Dufty and J. J. Brey, {\em unpublished}, available on {\tt
condmat/0201361}; J. W. Dufty, J. J. Brey and J. Lutsko, {\em
unpublished}, available on {\tt condmat/0201367}.

\bibitem{vannoije_fdt} 
I. Goldhirsch ad T. P. C. van Noije, {\em Phys. Rev. E} {\bf 61}, 3241
(2000).

\bibitem{lutsko} 
J. Lutsko, J. W. Dufty and J. J. Brey, {\em unpublished}, available on
{\tt condmat/0201369}.

\bibitem{kubo_1957} R. Kubo, {\em J. Phys. Soc. Jpn.} {\bf 12}, 570
(1957).

\bibitem{bkls} A. Barrat, J. Kurchan, V. Loreto and M. Sellitto, {\em
Phys. Rev. Lett.} {\bf 85}, 5034 (2000).

\bibitem{kurchan}
H. A. Makse and J. Kurchan, {\em Nature} {\bf 415}, 614 (2002).

\bibitem{tesi} Unpublished preliminar results, see A. Puglisi,
Ph.D. Thesis, available on
\verb<http://axtnt3.phys.uniroma1.it/~puglisi/thesis/<.

\bibitem{bird} G. A. Bird. {\em Molecular Gas Dynamics and the Direct
Simulation of Gas Flows}, Clarendon, Oxford, 1994

\bibitem{ciccotti} 
G. Ciccotti, G. Jacucci and I. R. McDonald, {\em J. Stat. Phys} {\bf
21}, 1 (1979)

\bibitem{jackson}
J. L. Jackson and P. Mazur, {\em Physica} {\bf 30}, 2295 (1964)

\bibitem{garzo} V. Garz\'o and J. Dufty, {\em Phys. Rev. E} {\bf 60} 5706
(1999).

\bibitem{Huilin} L. Huilin, L. Wenti, B. Rushan, Y. Lidan and
D. Gidaspow, {\em Physica A} {\bf 284}, 265 (1999).

\bibitem{losert} W. Losert, D.G.W. Cooper, J. Delour, A. Kudrolli and
J.P. Gollub, {\em Chaos} {\bf 9}, 682 (1999).

\bibitem{menon} K. Feitosa and N. Menon, preprint cond-mat/0111391.

\bibitem{Wildman} R.D. Wildman and D.J. Parker, {\em Phys. Rev. Lett.}
{\bf 88}, 064301 (2002).

\bibitem{MontaneroHCS} J. M. Montanero and V. Garz\'o, {\em
Gran. Matter} {\bf 4}, 17 (2002).

\bibitem{Clelland}
R. Clelland and C. M. Hrenya, Phys. Rev. E {\bf 65}, 031301 (2002). 

\bibitem{MontaneroShear} J. M. Montanero and V. Garz\'o, preprint
cond-mat/0201175.

\bibitem{puglisi2} U. Marini Bettolo Marconi and A. Puglisi, preprints
cond-mat/0112336 and cond-mat/0202267.

\bibitem{barrat_trizac} A. Barrat, E. Trizac, preprint
cond-mat/0202297-

\bibitem{martin} Ph. A. Martin and J. Piasecki, {\em Europhys. Lett.}
{\bf 46}, 613 (1999).


\end{thebibliography}
\end{document}